\documentclass[]{iopart}

\jl{6}   % Submit to CQG

\newcommand{\lie}{{\cal L}_N}
\newcommand{\tder}{\left( \partial_t - \lie \right)}

\begin{document}

\title{The constraints as evolution equations for numerical
  relativity}

\author{
  Adrian P Gentle\footnote[1]{Department of Mathematics,  University of Southern Indiana,
    Evansville, IN 47712-3596},
  Nathan D George\footnote[2]{DAMTP, University of Cambridge, Cambridge CB3 0WA, United
    Kingdom},
  Arkady Kheyfets\footnote[3]{Department of Mathematics, North Carolina State
    University, Raleigh, NC 27695-8205},
  and Warner A Miller\footnote[4]{Department of Physics, Florida
    Atlantic University, Boca Raton, FL 33431}}

\address{Theoretical Division, Los Alamos National Laboratory, Los
  Alamos, NM 87545}
\ead{apgentle@usi.edu}

%\date{\today}
\begin{abstract} 
  The Einstein equations have proven surprisingly difficult to solve
  numerically.  A standard diagnostic of the problems which plague the
  field is the failure of computational schemes to satisfy the
  constraints, which are known to be mathematically conserved by the
  evolution equations.  We describe a new approach to rewriting the
  constraints as first-order evolution equations, thereby guaranteeing
  that they are satisfied to a chosen accuracy by any discretization
  scheme.  This introduces a set of four subsidiary constraints which
  are far simpler than the standard constraint equations, and which
  should be more easily conserved in computational applications.  We
  explore the manner in which the momentum constraints are already
  incorporated in several existing formulations of the Einstein
  equations, and demonstrate the ease with which our new
  constraint-conserving approach can be incorporated into these
  schemes.
\end{abstract}

%\submitto{\CQG} 

\setcounter{footnote}{0}
%\maketitle

\section{Introduction}

Despite significant recent advances in both computational power and
algorithmic complexity, there remain significant unresolved problems
with numerical implementations of the Einstein equations.  Perhaps the
most exciting recent developments are the many new three-plus-one
dimensional formulations of these equations, which, at least in part,
provide greater stability than the original ADM formulation (see
\cite{lehner} for a review).

In particular, formulations of the Einstein equations in strongly
hyperbolic, flux-conservative form have opened the way for the
application of algorithms and techniques originally developed for
computational fluid dynamics (see, for example, \cite{bona95}).
Despite the great promise and significant advances, modern numerical
relativity codes are still unable to fully simulate the complete
coalescence of binary black hole systems.  A full understanding and
complete simulation of such systems are of vital importance to the
analysis of gravitational wave signals collected by LIGO and similar
gravitational wave detectors.

Until recently, almost all three-plus-one dimensional formulations of
the Einstein equations have shared one common feature --- the
constraint equations are monitored through evolution, but never again
solved once the initial data is constructed.  An exception to this is
the Texas group, who resolve the elliptic constraint equations after
every few timesteps \cite{texas}.  The four constraints are more
typically treated as additional conditions which are monitored while
the spacetime is evolved.  In fact the constraints, and the
Hamiltonian constraint in particular, have been found to be excellent
prognosticators for the accuracy and stability of the numerical
solution.  The rule of thumb appears to be that when the Hamiltonian
constraint ``explodes'', the code will crash shortly thereafter.  To
the best of our knowledge, all three-plus-one dimensional formulations
of the Einstein equations suffer from these fundamental problems.

In this paper we propose an approach to transforming all four
constraint equations into evolution equations for a new set of
dynamical variables --- essentially the conformal factor and it's
spatial derivatives, quantities which reduce to the Newtonian
potential and force in the weak-field limit.  Our formulation is based
on the standard York-Lichnerowicz conformal decomposition and split of
the extrinsic curvature into its trace and trace-free parts. In this
sense it is similar in spirit, if not in detail, to the standard
approach to the initial value problem \cite{york79} and several modern
formulations of the evolution equations \cite{BSSN0,BSSN1}.
  
The underlying motivation for the new formulation presented in this
paper is the belief that it is the violation of the constraint
equations, and resultant generation of spurious energy-momentum
sources, which lead to instabilities in numerical implementations of
Einstein's equations.  By solving the Hamiltonian and momentum
constraints directly, especially in regions of the domain where the
gravitational fields are strongly dynamic, we automatically prevent
the run-away errors which typically appear in the Hamiltonian
constraint.  By automatically satisfying the constraints, we guarantee
conservation of energy-momentum.

The formulation of the constraints described in this paper is closest
to that of Bona, Mass\'o \etal \cite{bona92,bona95,bona98,bona99}, who
use the momentum constraints as evolution equations.  In their early
papers the constraints were used purely to eliminate spatial
derivatives of the extrinsic curvature, thereby casting the equations
in strongly hyperbolic form \cite{bona95}.  Only later are the
momentum constraints explicitly described as evolution equations
\cite{bona99}, although they had always played that role in the
formulation.  More recently Bona \etal \cite{bona03} have
incorporated the constraints into the evolution system by expanding
the Einstein equations themselves, thereby maintaining general
covariance.  Our current approach follows the spirit of their earlier,
non-covariant formulations.

We show that the BSSN formulation \cite{BSSN0,BSSN1} shares important
features with the Bona-Mass\'o (BM) approach, in particular their
common treatment of the momentum constraints.  We demonstrate that the
BSSN equations incorporate the momentum constraints as evolution
equations, and we highlight similarities in the way the momentum
constraints are treated in the BM and BSSN approaches.  We then
develop our own formulation of the momentum constraints, as well as a
new approach to rewriting the Hamiltonian constraint as an evolution
equation.  The goal of this formulation is the automatic conservation
of energy-momentum which we achieve by directly solving the constraint
equations, thus guaranteeing that they are satisfied throughout the
evolution.

We proceed as follows.  In the next section we briefly discuss the
constrained nature of general relativity, followed by an outline of
the approach taken by Bona and Mass\'o in treating the momentum
constraints.  Turning to the BSSN approach, we highlight the
momentum-conserving properties of the algorithm, before developing our
own approach.  We rewrite both the Hamiltonian and momentum
constraints as evolution equations, and demonstrate how transparently
these equations can be incorporated into existing evolution
algorithms.  Finally, we consider both the advantages and the issues
raised by our results.

\section{Constraints in General Relativity}

The constraint equations in Einstein's 1915 geometric theory of
gravitation play a cornerstone role -- general relativity is a fully
constrained theory.  If the four constraints are satisfied at every
point over every possible spacelike hypersurface of a spacetime, then,
necessarily, the entire spacetime is a solution of all Einstein's
equations.  It is surprising that the constraints have not played a
more pivotal role in numerical relativity, given their central
importance in theoretical developments of the theory.

The four constraint equations per point in spacetime are
\begin{eqnarray}
  \label{eqn:superhamiltonian}
  {\cal H} &=&  R + (\Tr K)^2 - K_{ij} K^{ij}  -  2 \rho = 0   \\ 
  \label{eqn:supermomentum}
  {\cal H}_i & = &  \nabla_j K^j_{\ i} - \nabla_i \Tr K - S_i =0,
\end{eqnarray}
known as the Hamiltonian and momentum constraints, respectively.  The
extrinsic curvature $K_{ij}$ is defined as
\begin{equation}
  \label{eqn:k}
  K_{ij} = - \frac{1}{2 N} \left\{ \partial_t g_{ij} - 2 \nabla_{(i} N_{j)}
  \right\},
\end{equation}
with $\rho$ and $S_i$ representing the energy-momentum source terms.
In the remainder of this paper we work in vacuum ($\rho=S_i=0$),
although our analysis is equally valid in the presence of matter.
 
As it stands, these constraints relate the six components each of
$g_{ij}$ and $K_{ij}$, and represent an initial value problem which
must be solved to obtain consistent initial data.  Mathematically the
constraints are conserved by the evolution equations, implying that if
they are satisfied on one slice of a foliation, they will be satisfied
on all future slices.  However, it is well know that computational
implementations suffer from serious errors which propagate in one or
more of the constraints.  It is this problem which we hope to overcome
by recasting the constraints as evolution equations.

\section{Bona-Mass\'o treatment of the momentum constraints}

It is our goal to recast all four constraint equations as evolution
equations by introducing a set of new variables.  The goal is to
expand the system of evolution equations to include the traditional
constraints, which necessarily introduces a set of new ``subsidiary
constraints'' which must be monitored during evolution.  These
subsidiary constraints take the place of the original Hamiltonian and
momentum constraints.

Such an approach has already been taken, perhaps indirectly, by Bona
and Mass\'o \etal while developing hyperbolic formulations of the
evolution equations \cite{bona92,bona95,bona98,bona99}.  Using the
momentum constraints to ensure the hyperbolicity of the full set of
first-order, flux-conservative evolution equations \cite{bona95}, the
Bona-Mass\'o (BM) formulation introduces a new set of variables
\begin{equation}
  \label{eqn:bonaconst}
  V_i = \frac{1}{2}\, g^{jk} \left( \partial_i g_{jk} - \partial _j
  g_{ik} \right).
\end{equation}
Evolution equations for the $V_i$ can be obtained by differentiation
and simplification using the $\dot g_{ij}$ equation together with the
momentum constraints \cite{bona95}.  However, the same result can be
obtained directly from the momentum constraints themselves, which take
the form
\cite{bona99}:
\begin{equation}
\label{eqn:bm_momentum}
  \partial_t V_k   =    - 2 K^{ij} \left( d_{ijk} - \delta_{ik} d^l_{\ lj}
  \right) +  K^{ij} \left( d_{kij} - \delta_{ik} d_{jl}^{\ \ l}
  \right) 
\end{equation}
where we have taken $N=1, N^i=0$ for brevity, and 
\begin{equation}
  d_{kij} = \frac{1}{2}\, \partial_k g_{ij} 
\end{equation}
are the first-order flux variables required to reduce the spatial
order of the Einstein equations.  In this approach the definition of
$V_i$, equation (\ref{eqn:bonaconst}), is relegated to the status of a
constraint relating the independent variables $V_k$, $g_{ij}$ and
$d_{kij}$.  These are the subsidiary constraints in the Bona-Mass\'o
approach.

Bona and Mass\'o provide an elegant incorporation of the momentum
constraints into the set of dynamical equations, although the four new
subsidiary constraints (\ref{eqn:bonaconst}) are still of a moderately
complex form, with no obvious manner of enforcing them naturally
within an evolution scheme.   Before describing an approach to
incorporating the Hamiltonian constraint into the evolution equations,
we pause to consider the status of the constraint equations in the
BSSN formulation.

\section{Momentum conservation in the BSSN formulation}

As outlined above, Bona and Mass\'o have shown that the momentum
constraints can be rewritten as first-order evolution equations, which
take the form (\ref{eqn:bm_momentum}) when $N=1$ and $N^i=0$.  They
introduce the new variable $V_i$, defined by equation
(\ref{eqn:bonaconst}), which appears naturally in the momentum
constraints when one commutes the spatial derivatives of $K_{ij}$ with
the time derivatives of $g_{ij}$ implicit in the definition of
extrinsic curvature.  This at once removes troublesome spatial
derivatives, and explains why eliminating spatial derivatives in the
evolution equation for $V_i$ with the momentum constraints is in fact
equivalent to using the momentum constraints themselves as evolution
equations.

In apparent contrast, the BSSN formulation introduces the new
variables
\begin{equation}
  \label{eqn:bssnconst}
  \tilde \Gamma^i = - \partial_j \tilde g^{ij},
\end{equation}
where $\tilde g^{ij}$ is the contravariant conformal metric.  An
evolution equation is then obtained by differentiation and commutation
of spatial and temporal derivatives.  However, the BSSN formulation
uses precisely the same approach to eliminating spatial derivatives of
the conformal, trace-free portion of the extrinsic curvature $\tilde
A^{ij}$ as described by Bona and Mass\'o \cite{bona95}.  In doing so,
the BSSN evolution equation for $\tilde \Gamma^i$ becomes equivalent
to the evolution equation obtained directly from the momentum
constraints themselves, in precisely the same manner as the BM
formulation.

In fact, the BM variable $V_i$ and the BSSN $\tilde \Gamma^i$ are
closely related.  Under the conformal decomposition
\begin{equation}
  g_{ij} = e^{4\phi} \, \tilde g_{ij}  \qquad \mbox{with} \qquad
  \det(\tilde g_{ij}) = 1
\end{equation}
the BM variables become
\begin{equation}
  \label{eqn:vs}
  V_i = \tilde V_i + 4\, \partial_i \phi
\end{equation}
where $\tilde V_i$ is defined like $V_i$, but in terms of the
conformal metric $\tilde g_{ij}$.  We see that $V_i$ splits naturally
into a portion $\tilde V_k$ determined entirely by the conformal
geometry, and the remainder which depends on the scale factor.  A
similar relationship was obtained previously in the case of a static
conformal factor \cite{bona98}.  

Expanding the conformal portion of $V_i$, and using the constraint
$\det \tilde g =1$,  we find that 
\begin{equation}
  \tilde V_j =
  - \frac{1}{2}\, \tilde g_{ij}\, \tilde \Gamma^i,  
\end{equation}
clearly showing that the BSSN variable $\tilde \Gamma^i$ is just the
conformal part of the BM variable $V^i$.  With this
realization, it is straightforward to use the momentum constraints, in
the form of equation (\ref{eqn:bm_momentum}), to obtain an evolution
equation for $\tilde \Gamma^i$.  Not surprisingly, the resulting
equation is precisely the one used in the standard BSSN formulation to
evolve $\tilde \Gamma^i$.

In their extensive analysis and review of existing formulations of the
Einstein equations, Shinkai and Yoneda note that the advantages of the
BSSN system over the standard ADM formulation are due entirely to the
introduction of the $\tilde \Gamma^i$ variables, and the subsequent
elimination of spatial derivatives using the momentum constraint
\cite{shinkai}.  In other words, the sole advantage of the BSSN
approach is the {\sl use of the momentum constraints as evolution
  equations}.  The BSSN formulation uses the momentum constraints to
evolve the conformal portion of $V_i$, and introduces the constraints
(\ref{eqn:bssnconst}) in their place.

The relative success of the BSSN formulation provides strong
motivation for incorporating the Hamiltonian constraint into the set
of evolution equations.  We do this below, as well as proposing an
alternative formulation of the momentum constraints, with the
advantage of a set of extremely natural, and very simple, subsidiary
constraints.  As we shall see, our approach to the momentum
constraints is opposite to the BSSN choice, since we evolve that
portion of $V_i$ arising directly from the conformal factor.

\section{Energy conservation: the Hamiltonian constraint as evolution
  equation}
 
The key to rewriting the Hamiltonian constraint as an evolution
equation is performing a conformal decomposition on the three-metric.
Rewriting the extrinsic curvature in terms of its trace and trace-free
parts will allow us to use the Hamiltonian constraint to evolve the
scale factor.

We begin with a conformal decomposition of the three-metric,
\begin{equation}
  \label{eqn:conformal}
  g_{ij} = e^{4\phi}\, \tilde g_{ij}
\end{equation}
with $\det \tilde g_{ij} = 1$, and split the extrinsic curvature
into its trace and trace-free parts,
\begin{equation}
  \label{eqn:basek}
  K_{ij} = A_{ij} + \frac{1}{3}\,  g_{ij} \Tr K
\end{equation}
where $\Tr A = 0$.  This allows us to rewrite the Hamiltonian
constraint as
\begin{equation}
  \label{eqn:hamiltonian}
  {\cal H} = R - A_{ij} A^{ij} + \frac{2}{3}\, (\Tr K) ^2 - 2\rho,
\end{equation}
where $R$ is the full three-dimensional Ricci scalar calculated from
the physical three-metric $g_{ij}$.  It is therefore a function of
$\phi$ and $\tilde g_{ij}$, together with their spatial derivatives
\cite{york79}.  It is not a function of $\dot \phi$, and thus does
not play a vital role in the current analysis.  We could in principle
expand $R$, as York does, into terms containing spatial derivatives of
$\phi$ together with the conformal curvature $R(\tilde g)$.

Our aim is to obtain an evolution equation for the conformal factor
from the Hamiltonian constraint, which requires an understanding of
where $\dot \phi$ appears in the constraint.  Writing
\begin{equation}
  A_{ij} = \left( \delta^a_i\, \delta^b_j - \frac{1}{3}\, g_{ij}\, g^{ab} \right) K_{ab}
\end{equation}
and applying the conformal decomposition to the definition of
extrinsic curvature,  equation (\ref{eqn:k}), we have
\begin{equation}
  K_{ab} = - \frac{1}{2N} \left(  e^{4\phi} \partial_t \tilde g_{ab}
  - 2 \nabla_{(a} N_{b)} + 4 g_{ab} \dot \phi \right)
\end{equation}
and since 
\begin{equation}
  \left( \delta^a_i\, \delta^b_j - \frac{1}{3}\, g_{ij}\, g^{ab} \right)
  g_{ab} = 0,
\end{equation}
it is clear that $A_{ij}$ does not depend directly on the time
development of $\phi$.  The only functional dependence on $\dot
\phi$ in the Hamiltonian constraint is thus within the $\Tr K$ term.

Under the conformal decomposition (\ref{eqn:conformal}) the trace of
the extrinsic curvature becomes
\begin{equation}
  \label{eqn:trk}
  \Tr K = \frac{1}{N} \nabla_i N^i - \frac{6 \dot \phi}{N},
\end{equation}
and noting that the Lie derivative of the conformal factor along the
shift vector $N^i$ is given by
\begin{equation}
  \lie \phi =  N^k \partial_k \phi\, + \frac{1}{6}\, \partial_k
  N^k = \frac{1}{6} \nabla_k N^k,
\end{equation} 
the trace of the extrinsic curvature can be written as
\begin{equation}
  \label{eqn:TrK}
  \Tr K = - \frac{6}{N} \tder \phi.
\end{equation}
This equation is often used to directly evolve the conformal factor
\cite{BSSN1}.  Instead, we treat it as a definition of $\Tr K$ in
terms of the time development of the conformal factor, allowing us to
rewrite the Hamiltonian constraint as the evolution equation for
$\phi$.

The time development of the conformal factor only enters the
Hamiltonian constraint through the term quadratic in $\Tr K$.
Proceeding formally, we use equation (\ref{eqn:TrK}) to rewrite the
Hamiltonian constraint ${\cal H}=0$ as
\begin{equation}
  \label{eqn:sqrt_hamiltonian}
\tder \phi = \pm \frac{N}{6} \sqrt{ 
      \frac{3}{2} \left( A_{ij} A^{ij} - R + 2\rho \right) }.
\end{equation}
This is a problematic result.  The square root causes strife whenever
those terms within it approach zero (potential computational problems)
or become negative.  In computational tests we find that errors which
typically appear as violations of the Hamiltonian constraint when
using the ADM evolution equations do indeed reappear as problems
``under the square root''.

There are several ways of proceeding from the Hamiltonian constraint
(\ref{eqn:hamiltonian}).  In general there is no need to replace both
$\Tr K$ terms, since we wish to maintain linearity in the time
derivative of the conformal factor. We can therefore write
\begin{equation}
  \label{eqn:finalham}
  \tder \phi  = \frac{N}{4}\,
  \left( \frac{ R(g) - A_{ij} A^{ij} - 2\rho}{\Tr K}   \right),
\end{equation}
providing an evolution equation for the conformal factor which does
not suffer from the problems mentioned above.  However, problems can
still arise if the denominator is zero.  

We note that equation (\ref{eqn:finalham}) can be rewritten as an
evolution equation for a new variable
\begin{equation}
  \xi = \frac{\Tr K\, \phi}{N},
\end{equation}
with the advantage that the evolution equation itself is singularity
free when $\Tr K=0$.  Although this form may be more advantageous in
computations, it merely relocates the problem to the calculation of
$\phi$ once $\xi$ is known.  

Another alternative is to monitor $\Tr K$ within the code, replacing
equation (\ref{eqn:finalham}) with an alternative whenever the
absolute magnitude of $\Tr K$ falls below some threshold.  For
example, equation (\ref{eqn:TrK}) implies that 
\begin{equation}
  \tder \phi = 0
\end{equation}
whenever the trace of the extrinsic curvature is zero.  For the
remainder of this paper we will use equation (\ref{eqn:finalham}).

\section{Momentum conservation}

We now turn to the momentum constraints, rewritten in terms of the
conformal decomposition described in the previous section.  Our aim is
to find a new set of variables which can be evolved using the momentum
constraints, and which have a simpler subsidiary constraint structure
than the BM formulation, equation (\ref{eqn:bonaconst}).

The momentum constraints are
\begin{equation}
  {\cal H}_j = \partial_i K^i_{\ j} + \Gamma^i_{ik} K^{k}_{\ j}
- \Gamma^k_{ij} K^{i}_{\ k} - \partial_j \Tr K 
\end{equation}
into which we can substitute the definition of $K^i_{\ j}$ expressed
in terms of its trace and trace-free parts.  Combining the results in
the previous section we find that
\begin{equation}
  K_{ij} = A_{ij} + \frac{1}{3N}\,  g_{ij} \left( 
    \nabla_k N^k - 6 \dot \phi
  \right),
\end{equation}
and thus the momentum constraints are
\begin{eqnarray*}
  {\cal H}_j & = & \nabla_i A^i_{\ j} 
%+ \Gamma^i_{ik} A^{k}_{\ j}- \Gamma^k_{ij} A^{i}_{\ k} \\
%& &
 - \frac{2}{3}\partial_j \left( \frac{ \nabla_k N^k}{N}  \right) 
 + 4 \partial_j \left( \frac{\dot \phi}{N} \right) .
\end{eqnarray*}
The plan is to define the new variable
\begin{equation}
  \Phi_j = \partial_j \phi,
\end{equation}
and commute the temporal and spatial partial derivatives in the
momentum constraint to obtain an evolution equation for $\Phi_j$.  We
note that $\Phi_j$ can be viewed as that portion of Bona-Mass\'o's
$V_i$ variable which is derived purely from the scale factor, as shown
by equation (\ref{eqn:vs}).  This is opposite to the choice made in
the BSSN formulation, where the momentum constraints are used to
evolve the conformal portion of $V_i$.

Continuing in the manner outlined above, the momentum constraints
take the form
\begin{equation}
  \partial_t \Phi_j = 
   \frac{1}{6} \partial_j \nabla_k N^k 
  - \frac{1}{6}\, \Tr K\, \partial_j N
  - \frac{N}{4} \nabla_i A^{i}_{\ j},
\end{equation}
which provides an evolution equation for $\Phi_j$.  However, this can
be recast in the more convenient form 
\begin{equation}
  \label{eqn:momentum}
  \tder \Phi_j = 
   \frac{1}{6} \partial_j \partial_k N^k 
  - \frac{1}{6}\, \Tr K\, \partial_j N
  - \frac{N}{4} \nabla_i A^{i}_{\ j},
\end{equation}
by expanding the covariant derivative of the shift.  This is the
desired evolution equation, derived from the momentum constraints, for
the new variables $\Phi_j$.

The stability properties of a formulation involving this evolution
equation for $\Phi_j$ remain to be investigated.  The BSSN formulation
explicitly removes spatial derivatives of $A^{ij}$ (or its conformal
part) to improve stability.  However, we argue that it is the fact
that this procedure introduces the momentum constraint as an evolution
equation which improves the overall stability, not simply the removal
of spatial derivatives.

\section{Implementing the ``constrained evolution'' algorithm}

In the previous sections we described how the standard constraint
equations can be recast as evolution equations for the conformal
factor and its spatial derivatives.  It is not hard to see that these
new forms of the constraints can be easily ``bolted onto'' existing
formulations of the evolution equations.

We first consider the classic $\dot g$-$\dot K$, or ADM,
formulation of the Einstein equations \cite{mtw}.   Incorporating
equations (\ref{eqn:finalham}) and (\ref{eqn:momentum}) into the ADM
system,
\begin{eqnarray}
  \tder g_{ij} & = & - 2 N K_{ij} \\
  \tder K_{ij} & = & N \left( R_{ij} - 2 K_{il} K^l_{\ j} + \Tr K K_{ij}
    \right)  - \nabla_i \nabla_j N,
\end{eqnarray}
can be approached in a number of ways.  We could, for example,
introduce $A_{ij}$, $\Tr K$, $\phi$ and $\Phi_j$ as auxiliary
variables, continuing to treat the physical metric and extrinsic
curvature as the fundamental variables.  This is not the most
straightforward approach, and in general, it would seem advantageous
to explicitly construct a conformal, trace-free generalization of the
ADM equations.

This approach evolves $\phi$ and $\tilde g_{ij}$ in place of
$g_{ij}$, and replaces $K_{ij}$ with $\Tr K$ and $\tilde A_{ij}$.  
The resulting  equations are identical to the equivalent BSSN
equations, namely
\begin{eqnarray}
  \tder \tilde g_{ij} & = & - 2 N \tilde A_{ij} \\
  \tder \tilde A_{ij} & = & e^{-4\phi} \left[ 
    N R_{ij} - \nabla_i \nabla_j N \right]^{\mbox{TF}} \\
  & & +\ N \left( \Tr
  K\,\tilde A_{ij} - 2 \tilde A_{ik} \tilde A^{k}_{\ j} \right)
\end{eqnarray}
where $\tilde A_{ij} = e^{-4\phi} A_{ij}$ and ``TF'' denotes the
trace-free portion of the bracketed expression.   Various expressions
can be derived to evolve $\Tr K$, including the standard BSSN equation
\begin{equation}
  \tder \Tr K = N \left\{ \tilde A_{ij} \tilde A^{ij} + \frac{1}{3}
  \left( \Tr K \right)^2 \right\} - \nabla_i \nabla^i N  + 2 N \rho .
\end{equation}
The conformal factor $\phi$ and its spatial derivatives $\Phi_j$ are
then evolved using the Hamiltonian and momentum constraints, equations
(\ref{eqn:finalham}) and (\ref{eqn:momentum}):
\begin{eqnarray}
  \tder \phi & = & \frac{N}{4}\,
  \left( \frac{ R(g) - A_{ij} A^{ij} - 2\rho}{\Tr K}   \right) \\
  \tder \Phi_j & = & 
   \frac{1}{6} \partial_j \partial_k N^k 
  - \frac{1}{6}\, \Tr K\, \partial_j N
  - \frac{N}{4} \nabla_i A^{i}_{\ j}.
\end{eqnarray}

This formulation differs from the existing BSSN approach in two ways.
First, the momentum ``constraint'' is used to evolve the variables
$\Phi_j$ in place of $\tilde \Gamma^i$.  Second, the conformal factor
is evolved using the Hamiltonian ``constraint'', rather than being
driven directly by the trace of the extrinsic curvature through
equation (\ref{eqn:TrK}).

Similarly, an evolution algorithm based more directly on the BSSN
formulation \cite{BSSN0,BSSN1} can be obtained.  The standard BSSN
evolution equations are used to evolve $\tilde g_{ij}$, $\tilde
A_{ij}$ and $\Tr K$, as described above, together with the conformal
connection functions $\tilde \Gamma^i$ \cite{BSSN1}.  Since these
conformal connection functions are already evolved using the momentum
constraints in the BSSN formulation, our goal of using all four
``constraint equations'' as evolution equations can be achieved by
incorporating the Hamiltonian equations into the evolution scheme with
minor modifications.  

Our form of the Hamiltonian constraint, equation (\ref{eqn:finalham}),
is used to replace the standard BSSN evolution equation for the
conformal factor $\phi$, which is simply equation (\ref{eqn:TrK}).  In
this extended-BSSN approach the fundamental variables are still the
original set, namely $\phi, \tilde g_{ij}, \tilde A_{ij}, \Tr K$ and
$\tilde \Gamma^i$, with the only alteration being that the conformal
factor is evolved using (\ref{eqn:finalham}) rather than
(\ref{eqn:TrK}).  This fully-constrained approach is a trivial
addition to the standard BSSN formulation.

\section{Discussion}

By recasting the standard Hamiltonian and momentum constraints as
evolution equations, we have guaranteed that the evolution scheme
conserves energy-momentum.  We have also introduced the four
subsidiary constraint equations
\begin{eqnarray}
  \det \tilde g_{ij} &=&  1\\
  \partial_i \phi &=& \Phi_i 
\end{eqnarray}
in place of the original Hamiltonian and momentum constraints.  These
subsidiary constraints, or consistency conditions, ensure that the
derived variables $\Phi_j$ and $\phi$, which are evolved
independently, continue their expected relation to the original set of
physical variables.

The new subsidiary constraints are much simpler than the original
Hamiltonian and momentum constraints.  Codes which violate the
Hamiltonian and momentum constraints introduce, through computational
errors, spurious energy-momentum sources.  However, violation of the
subsidiary constraints does not result in the same physical problem,
since the original constraints (now in the form of evolution
equations) are still satisfied to machine precision.  Thus no spurious
energy-momentum can be injected into the system by purely
computational errors.  It seems likely that this advantage explains
much of the success of the BSSN formulation, in comparison with the
ADM form of the equations.  In this sense, the BSSN approach may be
viewed as a natural generalization of the ADM evolution equations to
incorporate the momentum constraints.

The major difficulty with the constraint-evolution equations we have
obtained is the potential for singularities to develop in the
Hamiltonian, arising from zeros in the denominator.  Although this
problem can be formally avoided by writing the equation in terms of a
new variable, the conformal factor must still be calculated from this
new variable.  As such, this approach merely removes singularities
from the evolution equations without wholly avoiding the problem. One
solution, at least for black hole spacetimes, would be to use
constant-crunch ($\Tr K = \mbox{constant} \ne 0 $) slicings, which
have been shown to provide potentially stable foliations of black hole
spacetimes \cite{gentle01}.  This is, however, not an entirely
satisfactory solution, and we continue to investigate this issue.

Work is currently under way to investigate the relative stability of
this new formulation of the Einstein equations.  In particular, we are
interested in the relative merits of the various forms of the momentum
equations (BSSN $\tilde \Gamma^i$, BM $V_i$, $\Phi_j$).  We are also
exploring the integration of the full set of constraints into the
Bona-Mass\'o and BSSN formalisms to determine their mathematical
structure, and performing computational tests of the new constraint
equations.

\section*{Acknowledgments} 

The authors wish to thank Richard Matzner, Matt Anderson, Ruth
Williams and Mark Gray for enlightening conversations during the
completion of this work, together with the Walnut Brewery for
providing inspiration.  We also acknowledge the LANL LDRD/ER program
for financial support.

% --- REFERENCES ----------------------------------------------------


\section*{References}

\begin{thebibliography}{99} 
  
\bibitem{lehner} L.~Lehner, {\em Class.Quant.Grav.} \textbf{18}
  (2001) R25-R86

\bibitem{texas} Matt Anderson, Richard Matzner, 
  Private Communication, 2003.

\bibitem{bona92}  C.~Bona, J.~Mass\'o
  {\em Phys.Rev.Lett.} \textbf{68} (1992) 1097

\bibitem{bona95}  C.~Bona, J.~Mass\'o, E.~Seidel, J.~Stela,
  {\em Phys.Rev.Lett.} \textbf{75} (1995) 600

\bibitem{bona98}  C.~Bona, J.~Mass\'o, E.~Seidel, P.~Walker, 1998
  {\it Preprint} gr-qc/9804052.
  
\bibitem{bona99} A.~Arbona, C.~Bona, J.~Mass\'o, J.~Stela, 
  {\em  Phys. Rev.}, \textbf{D60} 104014, 1999.

\bibitem{bona03} 
  C.~Bona, T.~Ledvinka, C.~Palenzuela, M.~Zacek,
  {\em  Phys. Rev.}, \textbf{D67} 104005, 2003.


\bibitem{york79} James W. York, ``Kinematics and Dynamics of General
  Relativity'', in {\em Sources of Gravitational Radiation}, ed. L.
  Smarr, Cambridge University Press, 1979.

\bibitem{BSSN0} M.~Shibata, T.~Nakamura,
  {\em  Phys. Rev.}, \textbf{D52} 5428, 1995.

\bibitem{BSSN1} T.~W.~Baumgarte, S.~L.~Shapiro,
  {\em  Phys. Rev.}, \textbf{D59} 024007, 1998.

\bibitem{shinkai} H.~Shinkai, G.~Yoneda,
  2002 {\it Preprint} gr-qc/0209111.

\bibitem{mtw} C.~W.~Misner, K.~S.~Thorne and J.~A.~Wheeler, 
  \textit{Gravitation} (1973) 

\bibitem{gentle01} A.~P.~Gentle, D.~E.~Holz, A.~Kheyfets, P.~Laguna,
  W.~A.~Miller and D.~M.~Shoemaker,
  {\em  Phys. Rev.}, \textbf{D63} 064024, 2001.

\end{thebibliography}
\end{document}